\documentclass[twocolumn,showpacs,preprintnumbers,amsmath,amssymb,nofootinbib]{revtex4}

\usepackage{here}
\usepackage{graphicx}
\usepackage{dcolumn}
\usepackage{bm}

\begin{document}


\title{Algorithm-Based Analysis of Collective Decoherence in Quantum Computation}

\author{Shoko Utsunomiya$^{1,2}$}
\author{Cyrus P. Master$^3$}%
\author{Yoshihisa Yamamoto$^{1,2,3}$}

\affiliation{
$^{1}$Department of Information and Communication Engineering, The University of Tokyo, Tokyo 113-8654, Japan \\
$^{2}$National Institute of Informatics, Tokyo 101-8430, Japan\\
$^{3}$E.L.Ginzton Laboratory, Stanford University, Stanford, CA
94305-4088
}%


\date{\today}

\begin{abstract}
The information in quantum computers is often stored in identical
two-level systems (spins or pseudo-spins) that are separated by a
distance shorter than the characteristic wavelength of a reservoir
which is responsible for decoherence. In such a case, the collective
spin-reservoir interaction, rather than an individual spin-reservoir
interaction, may determine the decoherence characteristics. We use
computational basis states, symmetrized spin states and spin
coherent states to study collective decoherence in the
implementation of various quantum algorithms. A simple method of
implementing quantum algorithms using stable subradiant states and
avoiding unstable Dicke's superradiant states and Schr\"odinger's
cat states is proposed.
\end{abstract}

\pacs{03.67Lx, 03.65Yz, 03.67.-a}
\maketitle
\section{Introduction}

In future quantum information processing systems, quantum bits
(qubits) of information are likely to be stored in identical
two-level systems (spins or equivalent spins) that are spatially
separated to simultaneously allow independent control of single
qubits and mutual interaction between qubits.  If a system of $n$
spins interacts with a reservoir whose characteristic wavelength is
longer than the inter-spin distance, the system interacts
collectively with the bath. For instance, electron-spin-based
quantum computers \cite{r_loss} have an inter-spin distance on the
order of 10 to 100 nm, which can be far shorter than the wavelength
of a coupled low temperature phonon reservoir.  Another example is a
nuclear-spin-based quantum computer using doped impurities
\cite{r_kane} or a crystal lattice \cite{r_yama,r_ladd} in which a
remote paramagnetic impurity interacts with a nuclear spin system by
a long-range dipolar coupling \cite{r_slih}.  If the distance
between qubits is much smaller than their separation from the
impurity, the interaction can also be treated as collective.

In theory, if a system-bath interaction exhibits a high degree of
symmetry, computation can be performed in a decoherence-free
subspace (DFS) that leaves the logical qubits immune to the
influence of the bath
\cite{r_kitaev,r_palma,r_duan,r_zana,r_zana2,r_lidar,r_knill,r_lidar2}.
However, if one restricts logical states to this subspace, it is
difficult to design simple gates to implement quantum algorithms.
Consider, by comparison, a standard model of quantum computation,
which labels logical states (computational basis states) by the
tensor-products of the two-level qubit states, and implements the
unitary evolution for a given algorithm by a sequence of
controlled-NOT (CNOT) gates and single-qubit rotations.  These
fundamental gates are physically convenient, as they can be created
simply through pairwise interactions between qubits and local
control pulses (i.e., by exploiting the innate tensor product
structure of the Hilbert space)\footnote{Note that we simplify
matters by neglecting fault-tolerant implementation and
system-environment interactions that break permutation symmetry.}.
By contrast, if logical states are defined within a DFS, the control
sequences are complicated by the requirement that the state does not
wander out of the DFS during evolution.  Thus, in the interest of
simple gate construction, we examine decoherence in the context of
the standard computational basis and the full Hilbert space.

Consider the unitary evolution of a quantum computer during the
course of a given algorithm.  Ideally, the state is described by a
trajectory in the $2^n$-dimensional complex Hilbert space.  If a
collective system-environment interaction is present, the
instantaneous decoherence rate can be parameterized by the
``proximity'' of the ideal trajectory to the decoherence-free
subspace, allowing us to infer how the decoherence rate evolves over
the course of the algorithm. Generally, the end result of a quantum
algorithm is encoded in the measurement of the final state; there is
thus some freedom to choose an initial state and/or trajectory that
minimizes the average proximity to the DFS.  In this fashion, one
can attempt to minimize the impact of decoherence due to the
collective system-reservoir interaction while maintaining the
simplicity of the tensor-product structure of the full Hilbert
space. As a compromise between a naive application of the standard
model and logical encoding in a DFS, this approach might be useful
to implement algorithms on a prototype quantum computer, where the
number of qubits is too modest to implement quantum error
correction.

In this paper, we introduce physical models for collective
spin-reservoir interactions, and examine relevant parameterizations
of the decoherence rate during the time evolution of specific useful
quantum algorithms. It is shown that judicious choice of the initial
state or gate sequence can often reduce the decoherence rate
parameter significantly.

A collective spin-reservoir interaction that is responsible for a
longitudinal relaxation process ($T_1$ process) can be modeled by
\begin{equation}
\hat{\mathcal{H}}_I = \sum_i\hbar k_i\left(\hat{a}_i^\dag\hat{J}^-
+\hat{a}_i\hat{J}^+\right). \label{equ.H_i_1}
\end{equation}
Here, $\hat{a}_i$ ($\hat{a}_i^\dag$) is an annihilation (creation)
operator of the bosonic reservoir and $\hat{J}^+$ ($\hat{J}^-$) is a
collective raising (lowering) operator of the spin system:
\begin{equation}
\hat{J}^{\pm} = \sum_{\ell=1}^n \hat{J}^{\pm}_\ell =
\frac{1}{2}\sum_{\ell=1}^n \sigma^\pm_\ell, \label{equ.H_i_2}
\end{equation}
where $\hat{\sigma}^\pm_\ell = \hat{\sigma}_\ell^x \pm
i\hat{\sigma}_\ell^y$ are Pauli operators for spin $\ell$.

In this model, the longitudinal relaxation time $T_1$ can be much
shorter than that for a single isolated spin; this phenomenon is
known as superradiant decay in atomic physics \cite{r_dicke}.  In
contrast, the so-called subradiant states have negligible $1/T_1$.
To characterize this longitudinal relaxation rate, it will be useful
to define simultaneous eigenstates of the squared total angular
momentum $\hat{J}^2$ and the z-component $\hat{J}^z$ of the total
angular momentum $\hat{J}$, notated symmetrized spin states, and
discussed further in Sec. \ref{sec.sym_basis}.  By projecting the
ideal trajectory onto the symmetrized states, we can calculate a
time-dependent parameter to estimate the $T_1$ relaxation rate.

The collective spin-reservoir interaction that we adopt to model
transverse relaxation ($T_2$) processes is
\begin{equation}
\mathcal{H}_I = \sum_i \hbar g_i \hat{a}^+_i\hat{a}_i\hat{J}^z.
\end{equation}
Here $\widehat{J}^z$ is the z-component of the total angular
momentum $\widehat{J}$,
\begin{eqnarray}
\widehat{J}^{z}=\sum^{N}_{j=1}\widehat{J}_{j}^z=
\frac{1}{2}\sum^{n}_{j=1}\widehat{\sigma}_{j}^z \ \ .
\end{eqnarray}

In order to gain a physical picture for the collective $T_1$ and
$T_2$ decoherence behaviors in an $N$ spin system, it is convenient
to use a spin coherent state which is a linear superposition of
symmetrized spin states with the same total angular momentum $j$ and
the third quantum number $\alpha $ \cite{r_boni,r_arre}. The
$Q(\alpha )$ representation of the spin state in terms of spin
coherent states can easily identify a macroscopically separated
linear superposition state which is subject to strong collective
decoherence.

We will examine for modest $n$ the collective decoherence properties
of two representative quantum algorithms: the Deutsch-Josza
algorithm \cite{r_dj} and quantum search \cite{r_grov}. We find a
few cases in which the system state must pass through vulnerable
states, such as Dicke's superradiant states and macroscopically
separated linear superposition states, and so is vulnerable to rapid
longitudinal relaxation and transverse relaxation due to large
instantaneous coupling to the external reservoir.  We propose that
simple modifications to the algorithms can avoid this problem.

\section{Computational Basis States, Symmetrized Spin States and
Spin Coherent States}

In this section, we introduce three sets of basis states. The first
set is a standard computational basis formed by the tensor product
of $n$ two-level systems, and is most natural in discussing
gate-by-gate implementations of various algorithms.  The second
basis set consists of symmetrized states that are convenient to
calculate collective longitudinal and transverse relaxation rates.
The third is a set of spin coherent states that is suitable to
discuss $Q(\theta , \varphi)$ representation.

\subsection{Computational basis states}

In accordance with the standard model describing the state of $n$
spin-1/2 qubits, computational basis states are defined as
simultaneous eigenstates of $\left\{\hat{J}_1^2, \hat{J}_2^2,\ldots,
\hat{J}_n^2; \hat{J}_1^z, \hat{J}_2^z, \ldots, \hat{J}_n^z\right\}$:

\begin{subequations}
\begin{align}
\widehat{J}_{i}^{2}| j_{1}, j_{2}, \dotsm , j_{n} ; m_{1}, m_{2}, \dotsm , m_{n} \rangle \hspace{70pt} \nonumber\\
\hspace{20pt} = j_{i}(j_{i}+1) | j_{1}, j_{2}, \dotsm , j_{n} ; m_{1}, m_{2}, \dotsm , m_{n} \rangle \\
\widehat{J}_{i}^z| j_{1}, j_{2}, \dotsm , j_{n} ; m_{1}, m_{2}, \dotsm , m_{n} \rangle  \hspace{70pt} \nonumber \\
\hspace{20pt} = m_{i} | j_{1}, j_{2}, \dotsm , j_{n} ; m_{1}, m_{2},
\dotsm , m_{n} \rangle \ ,
\end{align}
\end{subequations}
where $j_i = 1/2$ and $m_i = \pm 1/2$.  As all the $j_i$ values are
fixed for a collection of two-level spins, we omit them in denoting
a computational basis state.  Thus, the Hilbert space is spanned by
the $2^n$ basis states $\left| m_1, m_2, \ldots, m_n\right>$.

\subsection{Symmetrized spin states}\label{sec.sym_basis}

\subsubsection{Definition}
Symmetrized spin states are defined as simultaneous eigenstates of
$\left\{\hat{J}_1^2, \hat{J}_2^2,\ldots,\hat{J}_n^2; \hat{J}^2,
\hat{J}^z\right\}$:
\begin{subequations}
\begin{align}
\hat{J}_i^2\left|j_1,j_2,\ldots,j_n;j,m\right>
&= j_i(j_i+1)\left|j_1,j_2,\ldots,j_n;j,m\right> \\
\hat{J}^2\left|j_1,j_2,\ldots,j_n;j,m\right>
&= j(j+1)\left|j_1,j_2,\ldots,j_n;j,m\right>\\
\hat{J}^z\left|j_1,j_2,\ldots,j_n;j,m\right>
&=m\left|j_1,j_2,\ldots,j_n;j,m\right>
\end{align}\label{sym_definition}
\end{subequations}
We again omit the $j_i$ labels from the symmetrized states.  Note
that $j \in\{0, 1, 2, \ldots, n/2\}$ for $n$ even and $j \in \{1/2,
3/2, \ldots, n/2\}$ for $n$ odd, and $m \in \left\{-j,-j+1,\ldots,
j\right\}$. Generally, each symmetrized state
$\left|j,m,\alpha\right>$ is $d$-fold degenerate;
$\alpha\in\left\{1,2,\ldots,d\right\}$ encapsulates all other
quantum numbers required to distinguish each basis state.
Symmetrized spin states can be linearly expanded in terms of
computational basis states via Clebsch-Gordan coefficients, whose
properties are summarized in Appendix \ref{app1}.
\subsubsection{Collective longitudinal relaxation rate}

To gauge the impact of the longitudinal relaxation process described
by Eq. (\ref{equ.H_i_1}), one could solve the Heisenberg equation of
motion for the joint density matrix of the system and reservoir,
under the influence of the system and bath internal Hamiltonians,
any externally applied control fields, and the system-reservoir
coupling.  However, such an analysis is prohibitive due to the large
number of degrees of freedom.  To simplify, we view the
system-reservoir coupling as a perturbation that causes transitions
between system spin states.   In the absence of any relaxation
process (i.e., to zeroth order), the system evolves in a unitary
fashion according to the algorithm of interest.  The transition rate
$\gamma$ due to system-environment coupling is calculated in
Appendix \ref{app2}; this rate varies throughout the time evolution,
as it is dependent on the system state.  The transition rate serves
as a parameter to characterize the instantaneous longitudinal
relaxation rate.

For the system-bath interaction described by Eq. (\ref{equ.H_i_1}),
it is convenient to expand the system state in the symmetrized
basis, as the environment acts on the spins through the collective
raising and lowering operators $\hat{J}^\pm$.  The transition matrix
elements for these operators are
\begin{equation}
\left|\left<j,m\pm
1,\alpha\right|\hat{J}^{\pm}\left|j,m,\alpha\right>\right|^2 =
\left(j\mp m\right)\left(j\pm m + 1\right). \label{jupdown}
\end{equation}
It is assumed that the bath is cold and resides in the ground state;
the downward transition is thus relevant.  The factor in Eq.
(\ref{jupdown}) represents the amplification in the collective
relaxation rate relative to the transition rate $\gamma_0$ for a
single spin.  Note that for $n$ even, the states $\left|j=n/2,
m=0\right>$ and $\left|j=n/2,m=1\right>$ have the highest value of
$\gamma$:
\begin{equation}
\gamma = (j+m)(j-m+1)\gamma_0=\frac{n}{2}\left(\frac{n+1}{2}\right)
\gamma_0
\label{gammaeq}
\end{equation}
The longitudinal relaxation rate per spin (that is, $\gamma/n$) is
thus enhanced by a factor of approximately $n/4$ compared to that of
an isolated single spin. For a quantum register of 4000 spins, the
longitudinal relaxation rate would be increased by three orders of
magnitude.  This enhanced relaxation rate is known as superradiant
decay \cite{r_dicke}.  On the other hand, the states
$|j,-j,\alpha\rangle$ $(j=n/2-1, \dots)$ have zero longitudinal
downward relaxation rate. If a reservoir cannot provide enough
energy to the spin system to cause an upward transition, those
states are completely stable against a $\widehat{J}^-$ process and
are thus referred to as ``subradiant states''. Eq. (\ref{jupdown})
also shows that the collective longitudinal relaxation rate
decreases with the total angular momentum $j$. Since the downward
transition matrix element for $m=0$ scales as $j(j+1)$, the state
$|j=1,m=0\rangle $ has a relaxation rate more than six orders of
magnitude smaller than the superradiant state $|j=n/2=2000,m=0
\rangle $.

An arbitrary spin state $|\psi \rangle$ can be expanded as a linear
combination of an orthonormal complete set of symmetrized states,
\begin{equation}
|\psi\rangle = \sum_{j,m,\alpha} c_{j,m,\alpha}|j,m,\alpha\rangle \ .\label{arbitrarystate}
\end{equation}
The probability density of eigenvalues $j$ and $m$ for a given state
are
\begin{eqnarray}
P(j,m) &=& \sum_{\alpha}|c_{j,m,\alpha}|^2 \nonumber \\
&=& \sum_{\alpha}|\langle j,m,\alpha|\psi\rangle|^2 \nonumber \\
&=& \langle \psi|\sum_{\alpha}|j,m,\alpha\rangle\langle j,m,\alpha|\psi\rangle \ .
\end{eqnarray}\

Since the longitudinal relaxation rate is dependent on only $j$ and
$m$ but is independent of $\alpha$, the probability of finding a
given state $|\psi\rangle$ in the subspace designated by $(j,m)$ is
sufficient for our purpose of studying the longitudinal relaxation.

The overall normalized longitudinal relaxation rate for an arbitrary
state $|\psi \rangle$ is thus given by
\begin{eqnarray}
\frac{\gamma}{\gamma _{0}}=\sum_{j} \sum_{m}P(j,m)[j(j+1)-m(m-1)] \ .
\end{eqnarray}

\subsubsection{Collective transverse relaxation rate}\label{transverse_relaxation}

The collective transverse relaxation rate is calculated by a model
for a quantum nondemolition (QND) measurement of $\widehat{J}^z$,
which is summarized in Appendix \ref{app3}. The off-diagonal element
$|j,m,\alpha\rangle \langle j',m',\alpha '|$ in the density matrix
decays with a rate proportional to $|m-m'|$ by such a QND
measurement process.

Since the transverse relaxation rate is only dependent on the
quantum number $m$, we can expand an arbitrary $n$-spin state $|\psi
\rangle$ in terms of the computational basis states
\begin{eqnarray}
|\psi \rangle= \sum _{m_1} \sum _{m_2} \dotsi \sum _{m_n} c_{m_{1}
m_{2} \dotsi m_{n}} |m_{1} m_{2} \dotsi m_{n} \rangle \ .
\label{eq.B.13}
\end{eqnarray}
After the $T_2$ relaxation process, the (mixed) state density
operator is
\begin{eqnarray}
\widehat{\rho}&=&\sum_{m_1, \dotsm, m_n}\sum_{m'_1, \dotsm, m'_n} c_{m_1,\dotsm, m_n}
c^*_{m'_1, \dotsm, m'_n} \nonumber \\
&&\times e^{-\varGamma _0 |m-m'|t} |m_1, \dotsm, m_n\rangle \langle m'_1,
\dotsm, m'_n| \label{eq.B.14}
\end{eqnarray}
where $m=\sum_{i=1}^{n} m_i$, $m'=\sum^{n}_{i=1} m'_i$ and
$\varGamma _0$ is a $T_2$ relaxation rate for a single spin.

The fidelity of the mixed state $\rho$ compared to the initial pure
state $|\psi \rangle$ is adopted as a measure for the degradation of
the state :
\begin{eqnarray}
F=\sqrt{\langle \psi |\widehat{\rho}|\psi \rangle} \ . \label{eq.B.15}
\end{eqnarray}
If we substitute Eqs. (\ref{eq.B.13}) and (\ref{eq.B.14}) into Eq.
(\ref{eq.B.15}), we obtain
\begin{eqnarray}
F=\left(\sum^{2^n -1}_{i=0} \sum^{2^n -1}_{j=0}|c_i|^2|c_j|^2
e^{-\varGamma_{i,j}t}\right)^{\frac{1}{2}} \ ,
\end{eqnarray}
where $\varGamma _{i,j}=\varGamma _0 |m_i- m_j|$. We approximate
$e^{-\varGamma _{i,j}t}\simeq 1-\varGamma _{i,j}t$ for a short
evolution time $(t\ll 1/\varGamma_{i,j})$. If we use the relation
$\sum^{2^n -1}_{i=0} \sum^{2^n -1}_{j=0}|c_i|^2|c_j|^2=1$, which is
obtained from $\text{Tr}(\widehat{\rho})=1$, we have
\begin{eqnarray}
F&=&1-\frac{1}{2} \sum^{2^n -1}_{i=0} \sum^{2^n -1}_{j=0}|c_i|^2|c_j|^2
\varGamma _0 |m_i - m_j|t \nonumber \label{equ.B.17}\\
&\equiv&1-\frac{1}{2}\varGamma t.
\end{eqnarray}
The normalized $T_2$ relaxation rate for an $n$-spin system is then
given by
\begin{eqnarray}
\frac{\varGamma}{\varGamma _0}=\sum^{2^n -1}_{i=0} \sum^{2^n -1}_{j=0}|c_i|^2|c_j|^2
|m_i - m_j|.
\end{eqnarray}
Note that the states $|j=0, m=0,\alpha \rangle$, with a
$d=n!/[(\frac{n}{2}+1)!\frac{n}{2}!]$ fold degeneracy, form a
decoherence free subspace (DFS).

\subsection{Spin-coherent states}
\subsubsection{Definition}

Spin-coherent states $|\theta,\varphi\rangle$ are defined as
eigenstates of the following non-Hermitian operator \cite{r_glau}:
\begin{equation}
\left[\hat{J}_z \sin \theta + e^{i \varphi} \cos ^{2}
\left(\frac{\theta}{2}\right)\hat{J}_{-} -e^{i \varphi} \sin
\left(\frac{\theta}{2}\right) \hat{J}_{+}\right]
|\theta,\varphi\rangle =0 \ . \label{eqscs}
\end{equation}
Eq. (\ref{eqscs}) resembles the definition of a coherent state of
harmonic oscillator, $(\widehat{a}-\alpha)|\alpha\rangle=0$, where
$\alpha$ is a complex eigenvalue. Spin-coherent states in the
$j=n/2$ subspace can be mathematically constructed by the rotation
operator $\widehat{R}_{\theta,\varphi}$ acting on the ground state
$|j=n/2,m=-n/2\rangle$ \cite{r_arre}:
\begin{eqnarray}
\left|\theta,\varphi\right> &=&
\widehat{R}_{\theta,\varphi}\left|\frac{n}{2},-\frac{n}{2}\right>
\label{equ 2.16} \\
\widehat{R}_{\theta,\varphi} &=& e^{i\theta(\widehat{J}^x \sin\varphi -\widehat{J}^y \cos\varphi)}
\label{equ 2.17} \ .
\end{eqnarray}
As $\widehat{R}_{\theta,\varphi}$ commutes with $\hat{J}^2$,
$\left|\theta , \varphi \right>$ is an eigenstate of $\widehat{J}^2$
with the same eigenvalue $j(j+1)$ as $\left|n/2,-n/2\right>$.
Therefore, $\left|\theta,\varphi\right>$ can be expanded as a linear
combination of the completely symmetric states $\left|j=n/2,
m\right>$,
\begin{eqnarray}
\left|\theta,\varphi\right>
=\sum^{n/2}_{m=-n/2}\frac{\tau^{m+\frac{n}{2}}}{(1+|\tau|^2)^{n/2}}
\binom{n}{m+\frac{n}{2}}^{\frac{1}{2}}\left|\frac{n}{2},m \right>,
\end{eqnarray}
where $\tau =e^{-i\varphi}\tan\left({\frac{\theta}{2}}\right)$ .
\\

Spin-coherent states are not mutually orthogonal. The inner product
of two spin-coherent states is given by
\begin{equation}
|\langle \theta',\varphi'|\theta,\varphi\rangle|^2=\cos^{2n}(\Phi) \
,
\end{equation}
where $\Phi$ is the angle between $\left|\theta,\varphi\right>$ and
$\left|\theta',\varphi'\right>$, and is given by
\begin{equation}
\Phi = \cos^{-1} [
\cos\theta\cos\theta'+\sin\theta\sin\theta'\cos(\varphi-\varphi') ].
\end{equation}
Spin coherent states satisfy a completeness relation:
\begin{equation}
(n+1)\int
\frac{d\omega}{4\pi}|\theta,\varphi\rangle\langle\theta,\varphi|=
\sum_{m}\left|\frac{n}{2},m\right>\left<\frac{n}{2},m\right|=\hat{I}.
\end{equation}
As the coherent states span the $j=n/2$ subspace and are linearly
dependent, they form an over-complete set.

According to Eq. (\ref{equ 2.16}), spin coherent states are produced
by rotation of the ground state $\left|n/2,-n/2\right>$, which is a
product state without any entanglement.  The operator in Eq.
(\ref{equ 2.17}) can be broken into a product of rotation operators
for each individual spin:
\begin{eqnarray}
\widehat{R}_{\theta,\varphi}=e^{i\theta(\widehat{J}_{1}^x\sin\varphi-\widehat{J}_{1}^y\cos\varphi)}
\otimes e^{i\theta(\widehat{J}_{2}^x\sin\varphi-\widehat{J}_{2}^y\cos\varphi)}\otimes \nonumber \\
\dotsi \otimes
e^{i\theta(\widehat{J}_{n}^x\sin\varphi-\widehat{J}_{n}^y\cos\varphi)}.
\end{eqnarray}
The spin coherent states generated by this rotation operation are
also unentangled product states.
\subsubsection{$Q$ representation}\label{sec:quasi-prob}

We introduce the $Q$-representation, defined as the overlap between
a given state $\left|\psi\right>$ and a spin-coherent state
$\left|\theta,\varphi\right>$:
\begin{eqnarray}
Q(\theta,\varphi) &=& |\langle \theta,\varphi|\psi \rangle|^{2} \ .
\end{eqnarray}
Figure \ref{fig3} shows the $Q(\theta,\varphi)$ representation of
the following states:
\begin{itemize}
\item $\frac{1}{\sqrt 2}(\left|j=4,m=4\right>+\left|j=4,m=-4\right>)$,
\item $\frac{1}{\sqrt 2}(\left|j=4,m=3\right>+\left|j=4,m=-3\right>)$,
\item $\frac{1}{\sqrt 2}(\left|j=4,m=2\right>+\left|j=4,m=-2\right>)$,
\item $\frac{1}{\sqrt 2}(\left|j=4,m=1\right>+\left|j=4,m=-1\right>)$, and
\item $\left|j=4,m=0\right> $.
\end{itemize}
We see a non-classical oscillation in the x-y plane due to
interference between the superposed symmetric states.  As the
difference between $m$ values of the components increases, the
number of oscillations increases as well.  As discussed in Sec.
\ref{transverse_relaxation}, the states with rapid oscillations, and
thus widely separated $m$ values, are particularly suspectable to
$T_2$ relaxation.  In this fashion, the structure of the
$Q$-function provides a measure of the collective transverse
relaxation rate.

\subsubsection{Spin coherent states for all $j$ manifolds}
So far, we have discussed spin coherent states for the completely
symmetric subspace $(j=n/2)$ through Eq. (\ref{equ 2.16}). We can
define spin coherent states for arbitrary values of $(0\le j \le
n/2-1)$ by applying the rotation operator $\hat{R}_{\theta,\varphi}$
to the subradiant states $\left|j < n/2, m=-j, \alpha \right>$
\cite{r_schw}. In this way, the Hilbert space can be partitioned
into separate spheres with different radii $j$, as shown in Fig.
\ref{fig2}.


\section{Application to selected quantum algorithms}
In the previous section, we introduced three basis sets to aid in
characterizing the decoherence during the execution of
representative quantum algorithms.  Instances of two quantum
algorithms will be analyzed: the Deutsch-Jozsa algorithm \cite{r_dj}
and Grover's data search algorithm \cite{r_grov}.


\subsection{Deutsch-Josza algorithm}
\subsubsection{Standard implementation}
The Deutsch-Jozsa algorithm consists of the following four steps as
shown in Fig. \ref{fig4}.

\begin{description}
\item[Step1] : Preparation of an $n$-qubit control register and a one-qubit work register.
\begin{eqnarray}
|\psi_1\rangle = |0\rangle_c \otimes |1\rangle _w \nonumber
\end{eqnarray}

\item[Step2] : Hadamard transformation.
\begin{eqnarray}
    \underrightarrow{\ \ \ \widehat{H}\ \ \ } \ \ \ |\psi_2\rangle =
    \frac{1}{\sqrt{2^n}}\sum^{2^{n}-1}_{x=0}|x\rangle_c \otimes\frac{1}{\sqrt{2}}(|0\rangle_w-|1\rangle_w) \nonumber
\end{eqnarray}

\item[Step3] : $f$-controlled-NOT gate operation.
\begin{eqnarray}
    \underrightarrow{\ \ \widehat{U}_{fcN}\ \ } \ |\psi_3\rangle =
    \frac{1}{\sqrt{2^n}}\sum^{2^{n}-1}_{x=0}(-1)^{f(x)}
    |x\rangle_c\otimes\frac{1}{\sqrt{2}}(|0\rangle_w-|1\rangle_w) \nonumber
\end{eqnarray}

\item[Step4] : Hadamard transformation.
\begin{eqnarray}
    \underrightarrow{\ \ \widehat{H}\ \ \ } \ \ |\psi_4\rangle=\frac{1}{2^n}\sum^{2^{n}-1}_{y=0}
    \sum^{2^{n}-1}_{x=0}(-1)^{f(x)+x\cdot y}|y\rangle_c    \otimes |1\rangle _w \nonumber
\end{eqnarray}
\end{description}

The $f$-controlled-NOT gate operation could be performed by
expressing $f(x)$ in disjunctive-normal form and adding CNOT gates
with multiple control bits for each expression.  For example,
consider a balanced function $f(x)$ that acts on $n=8$ qubits,
yielding one if the parity of $x$ is odd and zero otherwise; let
$x_{n-1}, x_{n-2}, \ldots, x_0$ be the binary digits of $x$.  Let
$\bar{x}_i$ represent the logical complement of $x_i$; the symbols
$\vee$ and $\wedge$ are logical OR and AND, respectively.  The
function $f(x)$ can be expressed as
\begin{eqnarray}
f(x) = \left(x_1 \wedge x_2 \wedge x_3 \wedge x_4 \wedge \bar{x}_5
\wedge \bar{x}_6 \wedge \bar{x}_7 \wedge \bar{x}_8\right) \nonumber \\
\vee
\left(x_1 \wedge x_2 \wedge x_3 \wedge \bar{x}_4 \wedge x_5 \wedge
\bar{x}_6 \wedge \bar{x}_7 \wedge \bar{x}_8\right) \vee \ldots
\end{eqnarray}
Clearly, there will be 128 bracketed expressions for any balanced
function with $n=8$.  Each expression can be computed in sequence in
a quantum circuit by flipping the complemented bits\footnote{The
reader may note that a much simpler, computationally efficient
circuit exists for the example provided.  However, the decomposition
described above is perhaps more general, and it is applicable to any
balanced function.}, performing a CNOT gate with all $n$ qubits as
control bits and the work qubit as the target, and again flipping
the complemented bits. Here, we consider each of these 128 blocks as
a single computational step; along with the Hadamard
transformations, this leads to a total of 131 steps in the
algorithm, including the initial state\footnote{The consequences of
this arbitrary partitioning of the algorithm evolution are discussed
in Sec. \ref{sec:Conclusion}.}.

For the $f(x)$ described above, we have calculated the projection of
the state after each of these 131 steps onto the symmetric state
basis; here, we only show in Fig. \ref{fig5} the projections at the
selected steps indicated in Fig. \ref{fig4}.  The normalized
relaxation rates can then be calculated at these instants from the
projections. In Fig. \ref{fig6}, the projections onto the
spin-coherent states is shown.  The initial state $|\psi _1\rangle $
has $P(j,m)=\delta _{j,4} \delta _{m,-4}$ , which is transformed to
a spin coherent state $\left|\psi _2 \right> =\left|\theta =
\frac{\pi}{2}, \varphi =0\right> $ by Hadamard transformation. This
spin coherent state has nonzero $P(j,m)$ values only in the $j=n/2$
subspace and is peaked at $m=0$, so that the state is vulnerable to
superradiant decay. After implementation of half of the
$f$-controlled-NOT gates ($\left|\psi_{3-2} \right>$), the
$Q$-function exhibits a macroscopically-separated linear
superposition, as shown in Fig. \ref{fig6}.   The state
$|\psi_{3-3}\rangle$ immediately before the second Hadamard
transformation is again a spin coherent state with an opposite phase
$\left|\theta =\pi/2, \varphi =\pi \right>$ compared to the state
$\left|\psi_2\right>$.

Figure \ref{fig7}A shows the evolution of the normalized $T_1$
relaxation rate at each of the 131 computational steps in the
implementation of this algorithm.  The $T_1$ relaxation rate is
highest immediately after the first Hadamard gate and immediately
before the second Hadamard gate, where the corresponding states
$|\psi _2\rangle$ and $|\psi _{3-3}\rangle$ are spin-coherent states
with a superradiant relaxation rate $\gamma/\gamma_0=18$. This
relaxation rate should be compared to the overall $T_1$ relaxation
rate $\gamma/\gamma_0=4$ for the case that $n=8$ spins are in the
same spin-coherent state but interact individually with an external
reservoir.

The normalized $T_2$ relaxation rate in the implementation of this
algorithm jumps from zero for the ground state to ${\gamma}/{\gamma
_0}\simeq 1.51$ for the spin-coherent state, and stays at the same
rate for the remaining steps; this is due to the fact that the 128
$n$-control-bit CNOT gates used to implement $(-1)^{f(x)}$ only
cause phase shifts, and do not alter the probabilities $P_{m_1 m_2
\dotsm m_n}$.  Indeed, all functions $f(x)$ lead to the same $T_2$
relaxation rate in the intermediate 129 steps of the algorithm. This
$T_2$ relaxation rate should be compared to the overall $T_2$
relaxation rate ${\gamma}/{\gamma_0}=4$ for the case that the $n=8$
spins are in the same spin coherent state but interact individually
with an external reservoir. Note that a macroscopically separated
linear superposition state $\frac{1}{\sqrt{2}}(|++\dotsm +\rangle +
|--\dotsm -\rangle )$ has the same collective relaxation rate
${\gamma}/{\gamma _0}=4$.

One may infer that the Deutsch-Jozsa algorithm for this particular
$f(x)$ function is vulnerable to collective $T_1$ relaxation
processes, but is rather robust against collective $T_2$ relaxation
processes.

Next, let us consider a different $f(x)$ that takes value one when
the last four qubits are of odd parity, and zero otherwise.   In
this case, the final state of the algorithm is $\left|\psi
_4\right>=|++++----\rangle$.

Figure \ref{fig8} shows the evolution of $P(j,m)$ of the $n=8$ qubit
control register state. Even though the state $|\psi _2\rangle$
immediately after the first Hadamard transform is still subject to
superradiant decay, the state $|\psi_{3-3}\rangle$ immediately
before the second Hadamard transform and the final state $|\psi
_4\rangle$ have a large support on more stable subradiant states $|j
\le n/2, m=-j \rangle$.The decreased $T_1$ relaxation rate for this
case, compared to the previous case, is obvious in Fig. \ref{fig7}B.


\subsubsection{Improved implementation}
It is suggestive from the above examples that balanced functions
$f(x)$ which result in final states with $m\simeq 0$ have rather
stable $|\psi_3\rangle$ and $|\psi_4\rangle$ states against
collective $T_1$ and $T_2$ processes. Therefore, external
longitudinal and transverse relaxation occurs primarily after the
first Hadamard transformation.

This problem can be resolved by starting the Deutsch-Jozsa algorithm
with a state which has half up-spins and half down-spins. For
instance, consider the initial state $|----++++\rangle$ for the
$n=8$ case.  This choice of the initial state does not change the
fundamental structure of Deutsch-Jozsa algorithm at all; if the
final state is found to be identical to the initial state, $f(x)$ is
a constant function.  However, the modified initial state leads to a
dramatic improvement in the $T_1$ relaxation rate, as shown in Fig.
\ref{fig7}C,D.

\subsection{Grover's data search algorithm}
\subsubsection{Standard implementation}
One application of the Grover iterant in the database search
algorithm consists of the following four steps, as shown in Fig.
\ref{fig9} \cite{r_grov}.

\begin{description}

\item[Step1] : Initialization of $n$ qubit control register.
\begin{eqnarray}
|\psi_1\rangle = |0\rangle = |--- \dotsm -\rangle \nonumber
\end{eqnarray}

\item[Step2] : Hadamard transformation.
\begin{eqnarray}
    \underrightarrow{\ \ \ \widehat{H}\ \ \ } \ \ \ |\psi_2\rangle =
    \widehat{H}|0\rangle = \frac{1}{\sqrt{2^n}}\sum^{2^{n}-1}_{x=0}|x\rangle \nonumber
\end{eqnarray}

\item[Step3] : Phase flip of the target state $|\tau\rangle$.
\begin{eqnarray}
    \underrightarrow{\ \ \ \ \widehat{I}_{\tau}\ \ \ \ } \ \
    \ |\psi_3\rangle = (\widehat{I} - 2 |\tau \rangle \langle \tau |) \widehat{H} | 0 \rangle
    \hspace{30pt} \nonumber
\end{eqnarray}

\item[Step4] : Hadamard transform.
\begin{eqnarray}
    \underrightarrow{\ \ \widehat{H}\ \ \ } \ \ |\psi_4\rangle= \widehat{H}
    (\widehat{I}-2|\tau \rangle \langle \tau |)\widehat{H}
    |0\rangle \nonumber
\end{eqnarray}

\item[Step5] : Phase flip of the $| 0 \rangle $ component.
\begin{eqnarray}
    \underrightarrow{\ \ \ \ \widehat{I}_{0}\ \ \ } \ \ \ |\psi _5\rangle = -(\widehat{I} - 2|0\rangle
    \langle 0|) \widehat{H} (\widehat{I} -2|\tau \rangle \langle
    \tau | )\widehat{H} |0\rangle \nonumber
\end{eqnarray}
\end{description}

Defining the iterant $\widehat{Q}$ as $-\widehat{I}_0 \widehat{H}
\widehat{I}_{\tau} \widehat{H}$, the state $|\psi _5 \rangle$ can be
rewritten as
\begin{eqnarray}
|\psi _5\rangle &=& \widehat{Q}|\psi _1\rangle \nonumber \\
                &=& -(\widehat{I} - 2|0\rangle \langle 0|){\widehat{H}}^2|0\rangle+2(\widehat{I}-2|0\rangle \langle 0|)\widehat{H}|\tau \rangle \langle \tau | \widehat{H}|0\rangle \nonumber \\
                &=& (1-4|U_{\tau_0}|^2)|0\rangle + 2U_{\tau_0}\widehat{H}|\tau\rangle   ,\label{equ 3.3}
\end{eqnarray}

where $U_{\tau0}=\langle\tau | \widehat{H} |0\rangle $.\\

The same rotation operator $\widehat{Q}$ translates the state
$\widehat{H} |\tau\rangle$ to
\begin{eqnarray}
\widehat{Q}\widehat{H}|\tau\rangle &=& -\widehat{I_0}\widehat{H}\widehat{I_\tau}\widehat{H}\widehat{H}|\tau\rangle \nonumber \\
                                        &=& (\widehat{I} - 2 |0\rangle \langle 0 |)\widehat{H}|\tau \rangle \nonumber \\
                                        &=& \widehat{H} |\tau \rangle - 2 U_{\tau_0}^{\ast }    |0\rangle. \label{equ 3.4}
\end{eqnarray}

It is clear from Eqs. (\ref{equ 3.3}) and (\ref{equ 3.4}) that the
rotation operator $\widehat{Q}$ preserves the two-dimensional vector
space spanned by $|0\rangle$ and $\widehat{H}|\tau \rangle$. The
operator $\widehat{Q}$ rotates any linear superposition of
$|0\rangle $ and $\widehat{H} |\tau \rangle $ by an angle $\theta
\simeq 2|U_{\tau_ 0}|$ in this plane. The number of iterations that
are required to transform an initial state $|0\rangle $ to a target
state $\widehat{H}|\tau \rangle $ is therefore given by
\begin{equation}
l_{tot}\approx\frac{({{\pi}/{2}})}{\theta} = \frac{\pi}{4}
\sqrt{2^n},
\end{equation}
where $|U_{\tau_0}|= 1\sqrt{2^n}$ is used. The target state is
obtained with near unity probability by the Hadamard transformation
of the final state, $\widehat{H}(\widehat{H}|\tau \rangle)=|\tau
\rangle$.

Each iterant can be split into the phase flip operation of the
target state $\widehat{I}_{\tau}$ and the
inversion-about-the-average-probability-amplitude operation
$-\widehat{H}\widehat{I}_0\widehat{H}$ \cite{r_grov}.   We will
examine the system states after the inversion operation for the
$l^{\text{th}}$ iterant.

Fig. \ref{fig10} shows $P(j,m)$ for $n=8$ after $l=1,3,5,7,10$ and
$12$ iterations, for the case of $|\tau \rangle = |255\rangle =
|++++++++\rangle $. Fig. \ref{fig11} plots $Q(\theta,\varphi)$ for
the same states. As expected, both plots of $P(j,m)$ and
$Q(\theta,\varphi)$ indicate a monotonic decrease in the probability
of finding the initial state $\widehat{H}|0\rangle$ and a monotonic
increase in the probability of finding the target state. In this
example, both the initial state $\widehat{H}|0\rangle$ and the
target state $|\tau \rangle$ are in the completely symmetric
($j=n/2$) subspace, and so are all the intermediate states belong to
this subspace. Therefore, one would expect that the implementation
of the Grover algorithm for this target state is vulnerable to a
collective longitudinal relaxation process.

The solid line in the left panel of Fig. \ref{fig12} shows the
normalized $T_1$ relaxation rate at each of 50 selected evolution times,
corresponding to the initial and final states, and the state after
each gate in the 12 iterants.  The $T_1$ relaxation rate is highest
immediately after the first Hadamard gate and slowest immediately
after the second Hadamard gate. The oscillatory behavior per
iteration continues to the midpoint ($l=6$ Grover iterations in this
case). The high $T_1$ relaxation rate corresponds to the state close to
$\widehat{H}|0\rangle$ and the low $T_1$ relaxation rate corresponds to
the state close to $|0\rangle$. At the mid-point, the phase of this
oscillation is flipped. The high $T_1$ relaxation rate now corresponds to
the state close to $\widehat{H}|\tau = 255\rangle $ and the low
$T_1$ relaxation rate corresponds to the state close to $|255\rangle $.

In the left panel of Fig. \ref{fig13}, the solid line indicates the
normalized $T_2$ relaxation rate at the same 50 computational steps.  The
relaxation rate is increased after the first Hadamard gate since that state
 $\widehat{H}|0\rangle $ has a fairly large dispersion on the quantum number $m$, but is
decreased after the second Hadamard gate, where the state is
approximately $\sim |0\rangle - \frac{1}{8}\widehat{H}|255\rangle$
(unnormalized) which has a much smaller dispersion on $m$. The $T_2$
relaxation rate is further increased immediately after the third
Hadamard gate, where the state is approximately $\sim
\widehat{H}|0\rangle + \frac{1}{8}|255\rangle$. The $T_2$ relaxation
rate is high again immediately before the last Hadamard gate, where
the state is approximately $\sim \widehat{H}|255\rangle$ which has
the same dispersion on $m$ as the state $\widehat{H}|0\rangle$, and
is low again immediately before the second last Hadamard gate, where
the state is approximately $\sim \frac{1}{8}\widehat{H}|0\rangle +
|255\rangle$.

Evidently, this implementation of the search algorithm is vulnerable
to both longitudinal and transverse relaxation processes.

In Fig. \ref{fig14}, $P(j,m)$ for $n=8$ after $l=1,3,5,7,10$ and
$12$ iterations is plotted for a different instance of the search
algorithm, where $|\tau \rangle =|15\rangle = |----++++\rangle $ is
the target state. As shown in the last panel , the target state
$|\tau \rangle$ in this case has support largely in the subspaces
where $j\in\{2,1,0\}$. The state transforms from an unstable
superradiant state to a more stable subradiant state as the number
of Grover iteration increases.  The solid lines in the right panels
of Figs. \ref{fig12} and \ref{fig13} present the calculated $T_1$
and $T_2$ relaxation rates for this instance.

\subsubsection{Improved implementation}
The vulnerability of the Grover algorithm implementation to
collective longitudinal and transverse relaxation processes stems
from the initial spin-coherent state $\widehat{H}|0\rangle $. In
order to circumvent this weakness, we can start with an alternative
initial state such as $|\gamma = 15 \rangle =|----++++\rangle $. If
we replace a standard Grover iteration shown above with
\begin{eqnarray}
\widehat{Q} = -(\widehat{I} -2|\gamma \rangle \langle \gamma
|)\widehat{H} (\widehat{I} -2|\tau \rangle \langle \tau|)\widehat{H}
\ ,
\end{eqnarray}
this new rotation operator preserves the two dimensional vector
space spanned by $|\gamma \rangle$ and $\widehat{H}|\tau \rangle$.
$U_{\tau _{\gamma}}=\langle \tau | \widehat{H}| \gamma \rangle$ is
equal to $1/\sqrt{2^n}$ or $ -1\sqrt{2^n}$, depending on $|\gamma
\rangle$ and $|\tau \rangle$. The plus and minus signs correspond to
the opposite rotation directions of the initial vector by a Grover
iteration. Here an absolute rotation angle $\theta \simeq 2|U_{\tau
_{\gamma}}|\equiv 2|\langle \tau|\widehat{H}\gamma\rangle|$ is
identical to the previous rotation angle $2U_{\tau _0}=2\langle \tau
|\widehat{H}|0\rangle $. Therefore, the $\sim \sqrt{2^n}$ successive
operations of $\widehat{Q}$ rotates an initial state $|\gamma
\rangle$ to the final state $\widehat{H}|\tau \rangle$, from which
we can find the target state $|\tau \rangle$ by a single Hadamard
transformation. In this way we can avoid the unstable spin coherent
state $\widehat{H}|0\rangle $ from the implementation of Grover's
algorithm simply by using any computational basis state with $m=0$
as an initial state.

The improved $T_1$ and $T_2$ relaxation performance by this choice of an
initial state is demonstrated by the dashed lines in Fig.
\ref{fig12} and \ref{fig13}.


\section{Conclusion}\label{sec:Conclusion}

We have studied the collective decoherence properties of $n$ qubit
quantum registers under implementation of selected quantum
algorithms.  A perturbative treatment of longitudinal relaxation
implies that it is preferable to modify quantum algorithms to
maximize the evolution time spent in the vicinity of subradiant
states $|j,m=-j \rangle $, and avoid unstable superradiant states.
Here, we have shown that slight modifications of the initial states
of the Deutsch-Jozsa and quantum search algorithms are advantageous
in weighting more heavily the subradiant states, if error correcting
codes or more complicated logical encoding is prohibitive due to
limited resources. Ideally, one might use the $T_1$ transition rate
as a parameter to optimize the gate decompositions and initial
states used to implement a given algorithm.  The optimal strategy
whereby this can be done remains an open question.

In our approach, the transition rates were calculated only at
selected discrete evolution times in each algorithm.  In a
forthcoming paper, we will present a finer grain analysis of the
transition rates, where the algorithm is described in terms of
single and two-qubit gates, and the state between each gate is
considered in the overall transition rates.  In addition, an
argument will be espoused as to how such an analysis can approximate
a continuous average over the evolution generated by a
time-dependent Hamiltonian consisting of pairwise interactions and
external single-qubit pulses.

The work of S.U. is partially supported by CREST, JST and that of
C.P.M. and Y.Y is partially supported by NTT Basic Research
Laboratories, SORST, JST and AFOSR under the contract of
F49620-01-1-0556-P003-4.

\appendix
\section{ Clebsch-Gordan coefficients}\label{app1}

\begin{table}[htbp]
\begin{center}

\begin{eqnarray}
\ \ \ \ \ \ \left[
\begin{array}{c|c|c|c||c}
\hline \ \ \frac{3}{2} \ \ & \ \ \ \frac{3}{2} \ \ \ \ \ \frac{1}{2} \ \ \ \ \ \frac{1}{2} \ \ \ & \ \ \ \frac{3}{2} \ \ \ \ \ \frac{1}{2} \ \ \ \ \ \frac{1}{2} \ \ \ & \ \ \frac{3}{2} & j \nonumber \\
\hline \ \ \frac{3}{2} \ \ & \ \ \ \frac{3}{2} \ \ \ \ \ \frac{1}{2} \ \ \ \ \ \frac{1}{2} \ \ \ & -\frac{3}{2} \ -\frac{1}{2} \ -\frac{1}{2} & \ -\frac{3}{2} & m \nonumber \\
\hline
\end{array}\right]\ \ \ \ \ \ \ \ \ \
\end{eqnarray}
\begin{eqnarray}
\widehat{U}=
\left [
\begin{array}{c|ccc|ccc|c}
\ \ 1\ \  &  &  &  &  &  &  & \\
\hline
 & \frac{1}{\sqrt{3}} & -\frac{1}{\sqrt{6}} & \frac{1}{\sqrt{2}} &  &  &  & \\
 & \frac{1}{\sqrt{3}} & -\frac{1}{\sqrt{6}} & -\frac{1}{\sqrt{2}} &  &  &  & \\
 & \frac{1}{\sqrt{3}} & \frac{2}{\sqrt{6}} & 0 &  &  &  & \\
\hline
 &  &  &  & \frac{1}{\sqrt{3}} & -\frac{1}{\sqrt{6}} & \frac{1}{\sqrt{2}} & \\
 &  &  &  & \frac{1}{\sqrt{3}} & -\frac{1}{\sqrt{6}} & -\frac{1}{\sqrt{2}} & \\
 &  &  &  & \frac{1}{\sqrt{3}} & \frac{2}{\sqrt{6}} & 0 & \\
\hline
 &  &  &  &  &  &  &\ \ 1\ \  \\
\end{array} \nonumber
\right]
\begin{array}{c}
|+++\rangle\\
\hline
|++-\rangle\\
|+-+\rangle\\
|-++\rangle\\
\hline
|+--\rangle\\
|-+-\rangle\\
|--+\rangle\\
\hline
|---\rangle\\
\end{array}
\end{eqnarray}
\caption{A diagonalizing matrix $\widehat{U}$ defined by
(\ref{DIFINITIONdiagU}) for the $n=3$ spin case. The Clebsch-Gordan
coefficients for $(j,m)=(\frac{1}{2},-\frac{1}{2})$ states represent
the subradiant states.} \label{T1-diagonalizeU}
\end{center}
\end{table}

Since a symmetrized spin state $|j,m,\alpha \rangle $ satisfies the
eigenvalue relations in Eq.(\ref{sym_definition}) and the
orthonormality relation, $\langle
j',m',\alpha'|j,m,\alpha\rangle=\delta _{jj'}\delta _{mm'}\delta
_{\alpha\alpha'}$, the squared total angular momentum
$\widehat{J}^2$ is a diagonal matrix in this basis:
\begin{equation}
\langle j',m',\alpha'|\widehat{J}^2|j,m,\alpha \rangle =
j(j+1)\delta _{jj'}\delta _{mm'}\delta _{\alpha\alpha'}    .
\end{equation}
By inserting two identity operators $\widehat{I}=\sum_{m_{1}\dotsi
m_{n}} |m_1 m_2 \dotsi m_n\rangle \langle m_1 m_2 \dotsi m_n|$
before and after $\widehat{J}^2$ in the above equation, we obtain
\begin{eqnarray}
\sum_{m_1' \dotsi m_n'} \sum_{m_1 \dotsi m_n} \langle j',m',\alpha'|m_1',m_2',\dotsi,m_n' \rangle  \nonumber \\
\langle m_1',m_2',\dotsi,m_n'|\widehat{J}^2|m_1,m_2,\dotsi,m_n\rangle \nonumber \\
\langle m_1,m_2,\dotsi,m_n|j,m,\alpha\rangle =j(j+1) \delta
_{jj'}\delta _{mm'}\delta _{\alpha\alpha'} \ .
\end{eqnarray}
This equation suggests that the non-diagonal matrix $\langle
m_1',m_2',\dotsi,m'_N|\widehat{J}^2|m_1,m_2,\dotsi,m_N \rangle$ can
be transformed into the diagonal matrix $\langle
j',m',\alpha'|\widehat{J}^2|j,m,\alpha \rangle$ with eigenvalues
$j(j+1)$ by multiplying a unitary matrix $\widehat{U}$ and its
inverse $\widehat{U}^{-1}$ from right and left of the non-diagonal
matrix, where
\begin{align}
\widehat{U}=\sum_{m_1,\dotsi ,m_N} \langle m_1,m_2,\dotsi m_N|j,m,\alpha\rangle . \label{DIFINITIONdiagU}
\end{align}
Each column of this real unitary matrix provides the probability
amplitude $c_{m_{1},\dotsi ,m_n}$ for linear expansion of
$|j,m,\alpha \rangle $ in terms of $|m_1,m_2,\dotsi,m_n\rangle .$,
i.e.
\begin{equation}
|j,m, \alpha \rangle =\sum_{m_1 \dotsi m_n} c_{m_1 \dotsi m_n}
|m_1,m_2,\dotsi m_n \rangle .
\end{equation}
The coefficients $c_{m_1,\dotsi ,m_n}$ are the Clebsch-Gordan
coefficients \cite{r_jj}.

According to the above argument, the symmetrized states $|j,m,\alpha
\rangle$ are mathematically constructed by the following procedure.

\begin{enumerate}
\item Divide computational basis states into the groups with different
total longitudinal quantum number $m = \sum^{n}_{i=1}m_i;$ For
instance, $n=3$ spin states are split into four subspaces
$m=-\frac{3}{2},-\frac{1}{2},\frac{1}{2},\frac{3}{2}$ as shown in
the right column in Table \ref{T1-diagonalizeU}.
\item Calculate $\widehat{J}^2$ in terms of computational basis states.
Here, non-zero elements appear only in block diagonal sub-matrices
belonging to the same subspace designated by $m$.
\item Find the real unitary matrix $\widehat{U}$ which diagonalizes
$\widehat{J}^2$, where the diagonal elements of the diagonalized
matrix are identically equal to the eigenvalue $j(j+1)$.
\item The probability amplitudes of computational basis states for
constructing each symmetrized state (Clebsch-Gordan coefficients)
are given by each column of the real unitary matrix $\widehat{U}$.
The example for $n=4$ is listed in the left part of Table
\ref{T1-diagonalizeU}.
\end{enumerate}

The $n$-spin symmetrized states are grouped into different subspaces
by their eigenvalues $j$ and $m$ as shown in Table
\ref{T2-degeneratetable}. The first column with $j=\frac{n}{2}$
represents a set of completely symmetric states, which   is often
referred to as angular momentum eigenstates. These states have no
degeneracy, so the total number of the completely symmetric states
is $n+1$, corresponding to
$m=\frac{n}{2},\frac{n}{2}-1,\dotsi,-\frac{n}{2}+1,-\frac{n}{2}$.
The second column with $j=\frac{n}{2}-1$ and
$m=\frac{n}{2}-1,\dotsi,-\frac{n}{2}+1$ represents
$d={}_{n}C_{1}-{}_{n}C_{0}=(n-1)th$ fold degenerate states. The
total number of eigenstates in this subspace is
$(n-1)\times(n-1)=(n-1)^2$. The third column with $j=\frac{n}{2}-2$
and $m=\frac{n}{2}-2, \ \dotsi , -\frac{n}{2}+2$ represents
$d={}_{n}C_{2}-{}_{n}C_{1}=\frac{n}{2}(n-3)th$ fold degenerate
states. The total number of eigenstates in this subspace is
$(n-3)\times \frac{n}{2}(n-3)=\frac{n}{2}(n-3)^2$ The final $j$
manifold ends with either $j=\frac{1}{2}$ (if $n$ is an odd number)
or $j=0$ (if $n$ is an even number). In general the degeneracy of
the states $|j,m \rangle$ is given by \cite{r_dicke},
\begin{equation}
\frac{n!(2j+1)}{(\frac{n}{2}+j+1)!(\frac{n}{2}-j)!} \ ,
\end{equation}
which is independent of $m$.

\begin{table}
\begin{eqnarray}
        \begin{array}{ccc|cccc}
        & & & j=\frac{N}{2} & j=\frac{N}{2}-1 & j=\frac{N}{2}-2 & \dotsi \\
        & & & (d=1) & (d=N-1) & (d=\frac{N}{2}(N-3)) & \dotsi \\
        \hline
        m&=& \frac{N}{2} & \frac{\ }{\ \ \ \ \ } &  &  & \\
        m&=& \frac{N}{2}-1 & \frac{\ }{\ \ \ \ \ } & \frac{\ }{\ \ \ \ \ } &  &  \\
        m&=& \frac{N}{2}-2 & \frac{\ }{\ \ \ \ \ } & \frac{\ }{\ \ \ \ \ } & \frac{\ }{\ \ \ \ \ } &  \\
         & &  & \vdots & \vdots & \vdots &  \dotsi \\
        m&=& -\frac{N}{2}+2 & \frac{\ }{\ \ \ \ \ } & \frac{\ }{\ \ \ \ \ } & \frac{\ }{\ \ \ \ \ } &  \\
        m&=& -\frac{N}{2}+1 & \frac{\ }{\ \ \ \ \ } & \frac{\ }{\ \ \ \ \ } &  &  \\
        m&=& -\frac{N}{2} & \frac{\ }{\ \ \ \ \ } &  &  &  \\
        \end{array}
        \nonumber
\end{eqnarray}
\caption{A distribution of symmetrized states $|j,m,\alpha
\rangle$.} \label{T2-degeneratetable}
\end{table}

\section{Collective longitudinal $(T_1)$ relaxation rate}\label{Collective longitudinal $(T_1)$ relaxation rate}\label{app2}
The collective spin-reservoir interaction Hamiltonian, which is
responsible for a $T_1$ relaxation process, is given by Eq.
(\ref{equ.H_i_1}). We will solve the Schr\"{o}dinger equation,
\begin{eqnarray}
i\hbar\frac{d}{dt}|\psi (t)\rangle=(\widehat{\cal H}_{0}+\widehat{\cal H}_{I})|\psi (t)\rangle \ , \label{B1}
\end{eqnarray}
with an initial state
\begin{eqnarray}
|\psi (0)\rangle=|j,m,\alpha \rangle \otimes \Pi _i |0\rangle \ .
\end{eqnarray}
We assume that all bosonic modes with a continuous spectrum are in
ground states at $t=0$. $\widehat{\cal H}_{0}=\hbar \omega
\widehat{J}^z +\sum _{i}\hbar \omega _{i}
\widehat{a}_{i}^{\dagger}\widehat{a}_{i}$ is a free Hamiltonian for
the combined spin-boson system. A time dependent solution of the
Schr\"odinger equation is
\begin{eqnarray}
|\psi (t) \rangle = c_{m,0} (t) e^{-i\omega mt}|j,m,\alpha \rangle|0 \rangle + \nonumber \\
\sum _{i} c_{m-1,1_i}(t) e^{-i \omega (m-1)t-i \omega _{i} t}|j,m-1,\alpha \rangle|1_i \rangle \ , \label{B3}
\end{eqnarray}
where $|0\rangle =\Pi _{i}|0\rangle _{i}$ and $|1_{i} \rangle =|0\rangle _{1}\dotsm |0\rangle _{i-1}|1\rangle _{i}\dotsm$.

Substituting Eq. (\ref{B3}) into Eq. (\ref{B1}), we obtain
\begin{eqnarray}
&&i\hbar [\dot{c}_{m,0}(t)e^{-i\omega mt}|j,m,\alpha \rangle|0 \rangle\nonumber \\
&+& \sum _{i} \dot{c}_{m-1,1_i}(t) e^{-i \omega (m-1)t-i \omega _{i} t}|j,m-1,\alpha \rangle|1_i \rangle \nonumber \\
&=&\hbar\sum _{i} k_{i}[c_{m,0}(t) e^{i \omega mt}\sqrt{j(j+1)-m(m-1)}\nonumber \\
&\times& |j,m-1,\alpha \rangle |1_i \rangle + c_{m-1,1_i}(t)e^{-i \omega (m-1)t-i \omega _i t} \nonumber \\
&\times& \sqrt{j(j+1)-m(m-1)}|j,m,\alpha \rangle |0\rangle] \ .\label{B4}
\end{eqnarray}
Multiplying $\langle j,m,\alpha | \langle 0|$ and $\langle
j,m,\alpha |\langle 1_i | $ on both sides of Eq. (\ref{B4}), we have
the coupled mode equations for $c_{m,0}(t)$ and $c_{m-1,1_i}(t)$,
\begin{eqnarray}
\dot{c}_{m,0}(t)&=&-i\sum _{i}k_i \sqrt{j(j+1)-m(m-1)} \nonumber \\
&\times& e^{i(\omega -\omega _i)t} c_{m-1,1_i}(t) \ ,\label{B5}\\
\dot{c}_{m-1,1_i}(t)&=&-i k_i \sqrt{j(j+1)-m(m-1)} \nonumber \\
&\times& e^{i(\omega -\omega _i)t} c_{m,0}(t) \ ,\label{B6}
\end{eqnarray}
The initial condition for the above equations is
\begin{eqnarray}
c_{m,0}(0)=1.\nonumber \\ \label {B7}
c_{m-1,1_i}(0)=0.
\end{eqnarray}
We can integrate Eq. (\ref{B6}), taking initial condition Eq.
(\ref{B7}) into account, and substitute $c_{m-1,1_i}(t)$ into Eq.
(\ref{B5}) to obtain the following integro-differential equation
\begin{eqnarray}
\dot{c}_{m,0}(t)=-\sum_{i} k_i^2[j(j+1)-m(m-1)]\nonumber \\
\int _{0}^{t}dt'e^{i(\omega - \omega _{i})(t-t')}c_{m,0}(t')dt' \label{B8}
\end{eqnarray}
If we replace $c_{m,0}(t')$ in the righthand side of Eq. (\ref{B8})
by $c_{m,0}(t)$ and replace the summation $\sum_{i}$ by an integral
$\int \rho (\omega _i) d\omega _i$ with an energy density of states
$\rho(\omega_{i})$, (\ref{B8}) is reduced to \cite{e_cohen}
\begin{eqnarray}
\dot{c}_{m,0}(t)=-\frac{\gamma}{2}c_{m,0}(t) \ ,
\end{eqnarray}
where
\begin{eqnarray}
\gamma=\frac{2\pi}{\hbar}k^2(\omega)[j(j+1)-m(m-1)]\rho(\omega) \ ,
\end{eqnarray}
where we neglect the frequency shift. Since
$\gamma_0=\frac{2\pi}{\hbar}k^2 \rho(\omega)$ is a longitudinal
relaxation rate for a single spin, the collective relaxation rate is
enhanced or suppressed by a factor of $j(j+1)-m(m-1)$.

\section{Collective transverse $(T_2)$ relaxation rate}\label{app3}
The collective spin-probe interaction Hamiltonian, by which a QND
measurement of $\widehat{J}^z$ is realized, is given by Eq.
(\ref{equ.H_i_1}). We assume the use of a single bosonic mode as a
quantum probe and thus the summation over mode index $i$ is
suppressed. We will solve the Schr\"odinger equation
\begin{eqnarray}
i\hbar \frac{d}{dt}|\psi (t)\rangle = \widehat{\cal H}_I |\psi (t) \rangle \ ,
\end{eqnarray}
where a free Hamiltonian $\widehat{\cal H} _0$ is suppressed. The
unitary time evolution operator in this case is reduced to
\begin{eqnarray}
\widehat{U}(t)=\exp [-igt \widehat{J}^z \widehat{a}^{\dagger}\widehat{a}] \ .
\end{eqnarray}
We assume the initial state for the spin and probe systems is given
by
\begin{align}
|\psi (0)\rangle =&[c_m |j,m,\alpha \rangle + c_{m'}|j',m',\alpha ' \rangle]_s\nonumber\\
&\otimes \sum_{n}c_n |n\rangle _p.
\end{align}
The state after the interaction is obtained by
\begin{eqnarray}
|\psi (t)\rangle &=&\widehat{U}(t)|\psi(0)\rangle \nonumber \\
&=& \sum _n[c_m e^{igmnt}|j,m,\alpha \rangle _s \label{C4} \\
&+& c_{m'} e^{igm'nt}|j',m',\alpha '\rangle _s]\otimes c_m |n\rangle _p , \nonumber
\end{eqnarray}
where $\widehat{J}^z|j,m,\alpha \rangle$ and $\widehat{a}^{\dagger}\widehat{a}|n\rangle =n|n\rangle$ are used.

The reduced density operator $\widehat{\rho} _s ^{(red)}$ for the
spin system corresponding to Eq. (\ref{C4}) has the following
non-zero components:
\begin{align}
\widehat{\rho} _s^{(red)} &\equiv \text{Tr}_{\rho}(|\psi\rangle \langle\psi|)  \label{C5}\\
&=[|c_m|^2|j,m,\alpha\rangle \langle j,m,\alpha|\nonumber\\
&\ +|c_{m'}|^2|j',m',\alpha'\rangle \langle j',m',\alpha '| \nonumber \\
&\ +c_m c_{m'}^{*} \sum _n |c_n|^2 e^{-ig(m-m')nt}|j,m,\alpha\rangle \langle j',m',\alpha ' | \nonumber \\
&\ +c_{m'} c_{m}^{*} \sum _n |c_n|^2 e^{-ig(m'-m)nt}|j',m',\alpha ' \rangle \langle j,m,\alpha |]
\nonumber
\end{align}
Eq. (\ref{C5}) suggests that the off-diagonal element has an
oscillation frequency broadening $\varDelta \omega = g|m-m'|\langle
\varDelta \widehat{n}^2 \rangle ^{1/2}$, where $\langle \varDelta
\widehat{n}^2 \rangle ^{1/2}=\sum _{n}|c_n|^2 (\widehat{n}-\langle
\widehat{n} \rangle)^2$ is the variance of the particle number. If
we consider this frequency broading as a continuous variable, the
off-diagonal elements $c_m c_{m'}^{*}$ and $c_{m'} c_{m}^{*}$ relax
with a rate given by
\begin{eqnarray}
\varGamma = cg|m-m'|\langle \varDelta \widehat{n}^2 \rangle ^{\frac{1}{2}} \ , \label{C6}
\end{eqnarray}
where $c$ is a constant determined by the distribution $|c_m|^2$.
For a single isolated spin in a linear superposition state where
$|m-m'|=1$, the transverse relaxation rate is $\varGamma _0 =cg
\langle \Delta\widehat{n}^2 \rangle ^{\frac{1}{2}}$. Therefore, the
collective transverse relaxation rate is either or suppressed by a
factor of $|m-m'|$, compared to that of a single spin.

\newpage
\bibliography{collective_dec_final}


\begin{widetext}

\begin{figure}
\begin{center}
\includegraphics[width=3in]{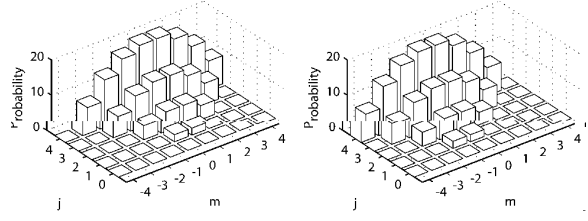}
\caption{Normalized matrix elements for upward (left) and downward
(right) transitions as a function of $j$ and $m$.} \label{fig1}
\end{center}
\end{figure}
\begin{figure}
\begin{center}
\includegraphics[width=3in]{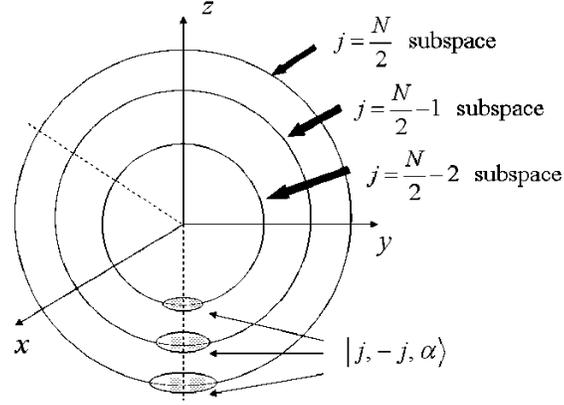}
\end{center}
\caption{Spin coherent states for arbitrary $j$ manifolds.}
\label{fig2}
\end{figure}

\begin{figure}
\begin{center}
\includegraphics[width=6in]{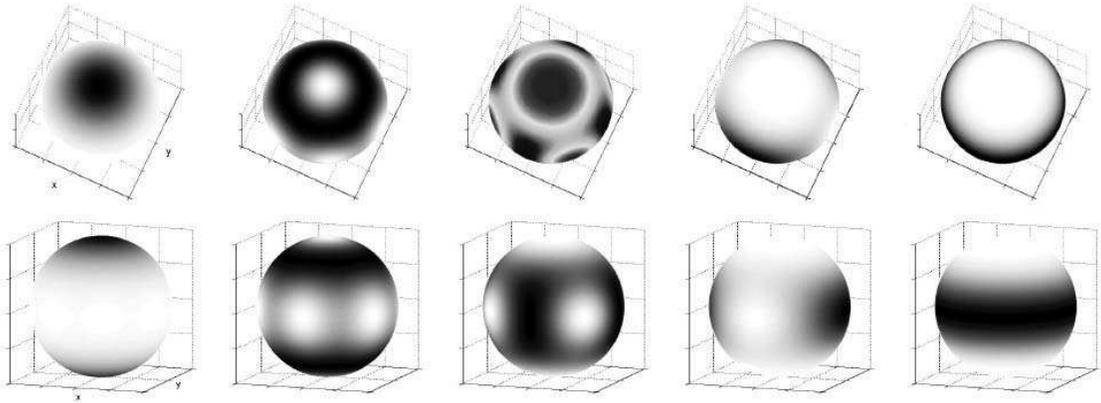}
\caption{ $Q(\theta, \varphi )$ representation of five states seen
from the ``north-pole'' (top) and from the equatorial direction
(bottom). $Q=1$ appears as black, while $Q=0$ appears as white.  The
five states (unnormalized) from left to right are
$\left|j=4,m=4\right>+\left|j=4,m=-4\right>$,
$\left|j=4,m=3\right>+\left|j=4,m=-3\right>$,
$\left|j=4,m=2\right>+\left|j=4,m=-2\right>$,
$\left|j=4,m=1\right>+\left|j=4,m=-1\right>$ and
$\left|j=4,m=0\right>$.}\label{fig3}
\end{center}
\end{figure}

\newpage
\begin{figure}
\begin{center}
\includegraphics[width=4in]{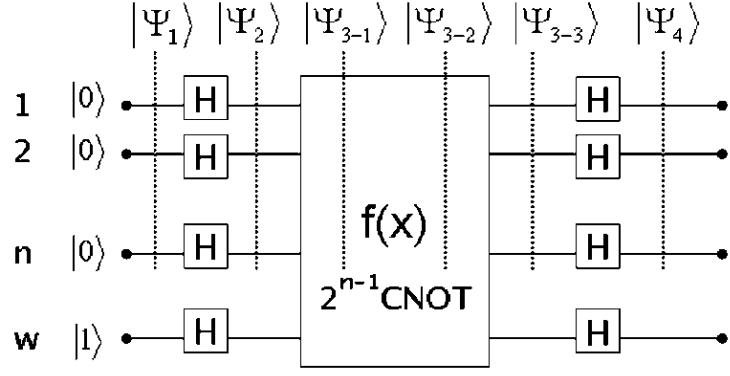}
\end{center}
\caption{Circuit for the Deutsch-Jozsa algorithm.} \label{fig4}
\end{figure}

\begin{figure}
\includegraphics[width=4in]{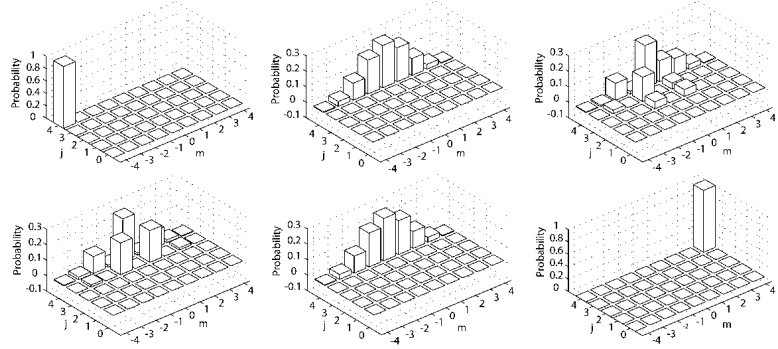}
\caption{Projections $P(j,m)$ onto the symmetrized states for the
Deutsch-Jozsa algorithm; $n=8$ and $f(x)$ is one iff the parity of
$x$ is odd. The plots, proceeding left-to-right in the first and
second rows, correspond to the projections of $\left|\psi _1
\right>$ (top left) through $\left|\psi _4\right>$ (bottom right),
indicated in Fig. \ref{fig4}. Note that the initial and final states
for this example are $\left|\psi_1\right>=\left|0\right>$ and
$\left|\psi_4\right>=\left|2^n -1\right>$.} \label{fig5}
\end{figure}

\begin{figure}
\includegraphics[width=5in]{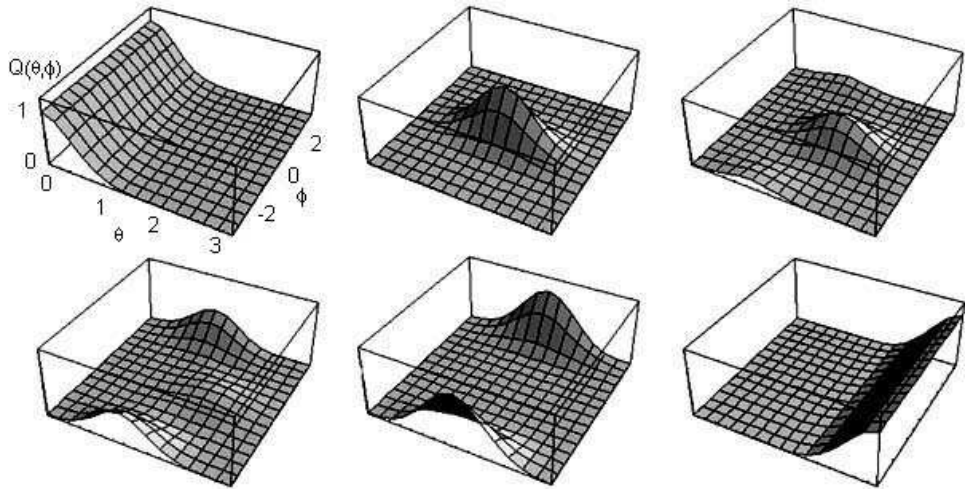}
\caption{$Q(\theta , \varphi)$ for the Deutsch-Jozsa
algorithm; the plots correspond to the states described in Fig.
\ref{fig5}. Note that in each graph, the unit sphere has been mapped
onto the $x$-$y$ plane; the transverse axes vary between $0\le
\theta \le \pi$ and $-\pi \le \varphi \le \pi$.} \label{fig6}
\end{figure}


\newpage
\begin{figure}
\begin{center}
\includegraphics[width=6in]{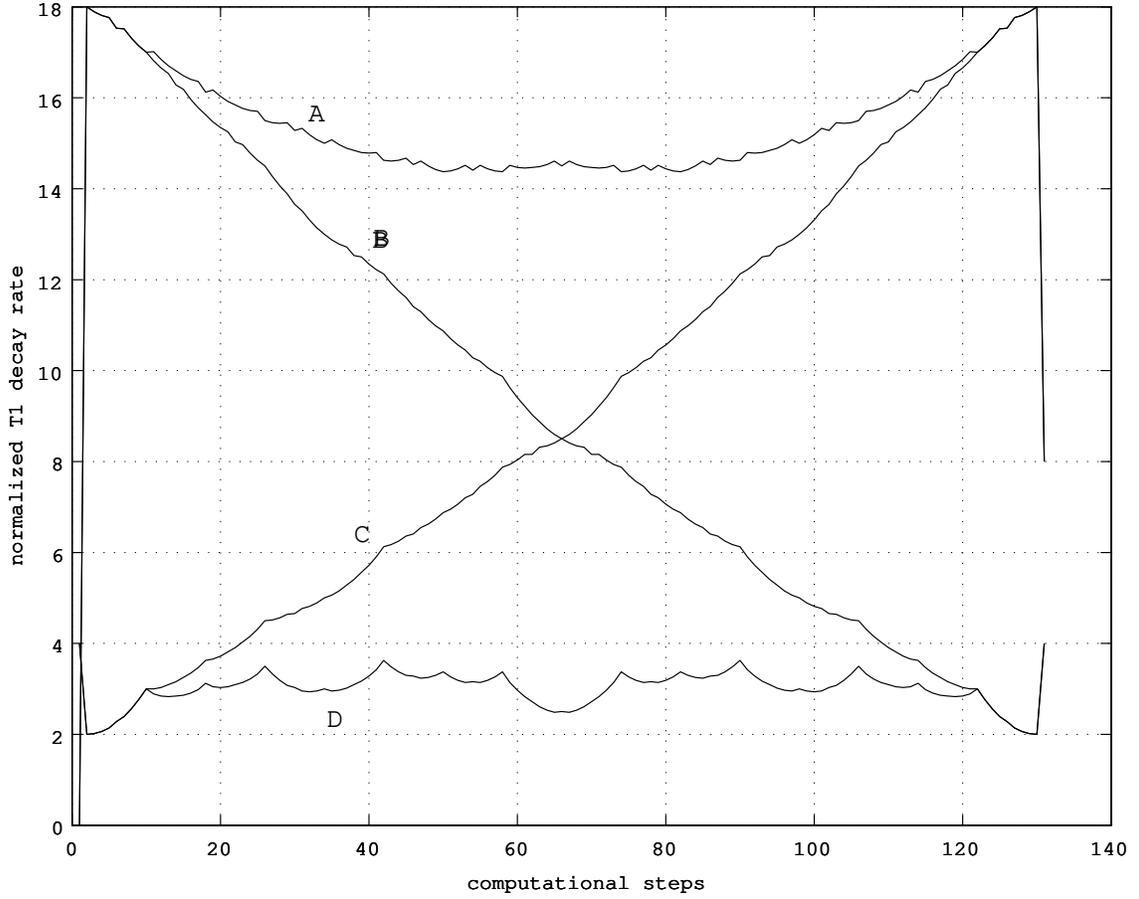}
\end{center}
\caption{The normalized $T_1$ relaxation rates at each of the 131
computational steps in the Deutsch-Jozsa algorithm. The traces
correspond to: (A) Unmodified algorithm, $f(x)=1$ if the parity of
$x$ is even. (B) Unmodified algorithm, $f(x)=1$ if the parity of the
last four bits of $x$ is even. (C) Improved algorithm, $f(x)$ as in
A. (D) Improved algorithm, $f(x)$ as in B.}
\label{fig7}
\end{figure}

\begin{figure}
\includegraphics[width=4in]{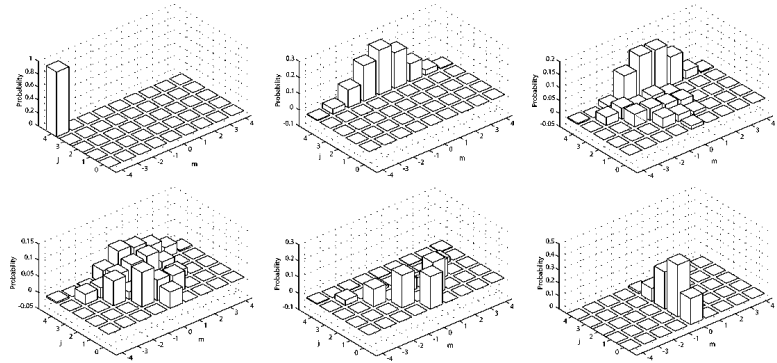}
\caption{$P(j,m)$ for states $\left|\psi _{1}\right>$ through
$\left|\psi _{4}\right>$ in Deutsch-Jozsa algorithm, $n=8$, $f(x)$
equalling one if the parity of the last four qubits is odd. Note
that the initial and final state are $\left|\psi _1\right> =
\left|0\right>$ and $\left|\psi _4\right> = \left|15\right>$. }
\label{fig8}
\end{figure}

\newpage
\begin{figure}
\begin{center}
\includegraphics[width=4in]{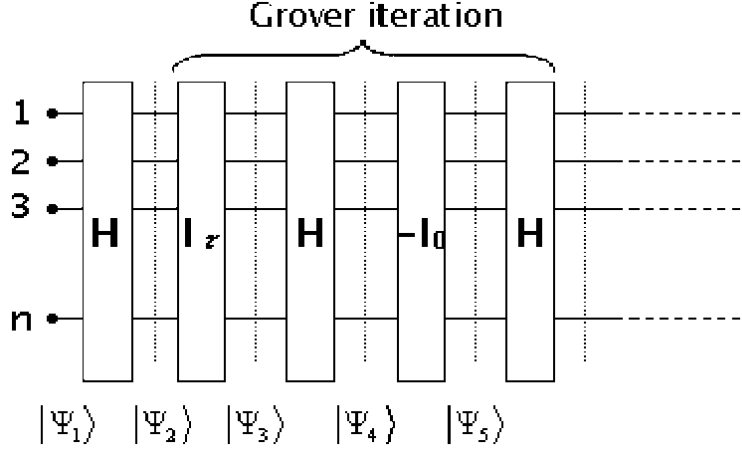}
\end{center}
\caption{Circuit topology for Grover's database search algorithm}
\label{fig9}
\end{figure}

\begin{figure}
\includegraphics[width=4in]{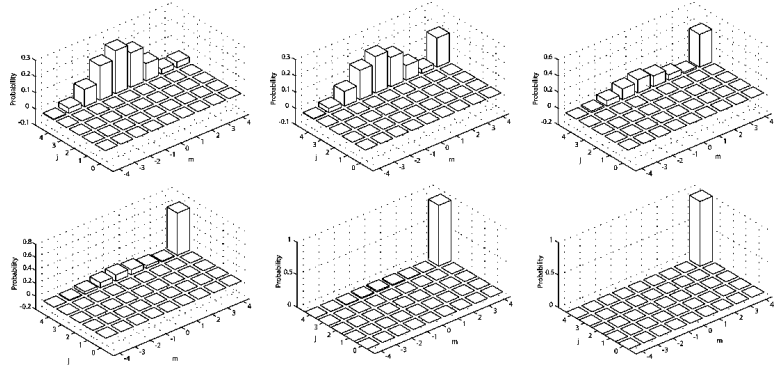}
\caption{ $P(j,m)$ of the control register states in the $n=8$
database search algorithm. The plots correspond to the states after
Grover iteration $\ell\in\{1,3,5,7,10,12\}$; $\ell=1$ is top left
and $\ell=12$ is bottom-right. The initial and final states are
$|\psi _1\rangle=|0\rangle $ and $|\psi _4\rangle=|2^n -1\rangle $.}
\label{fig10}
\end{figure}

\begin{figure}
\includegraphics[width=5in]{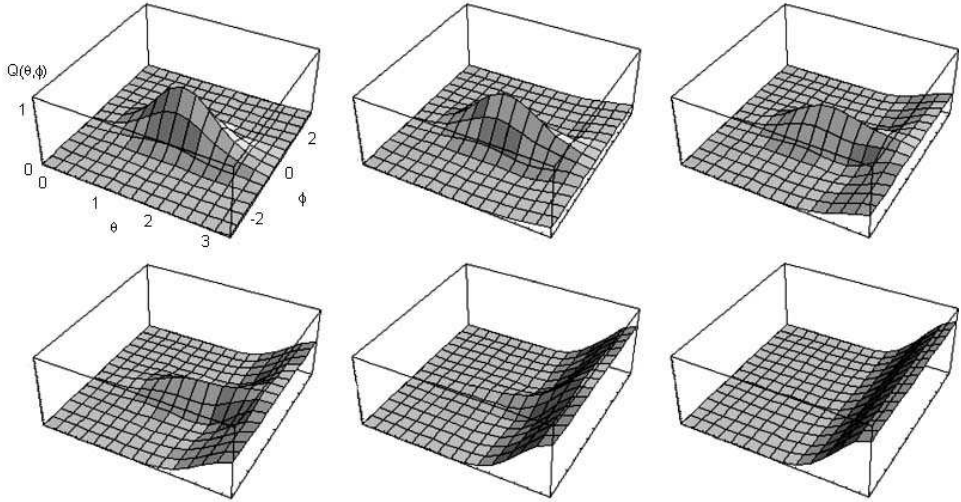}
\caption{ $Q(\theta,\varphi)$ of the control register states in the
$n=8$ database search algorithm. The plots correspond to the states
after Grover iteration $\ell\in\{1,3,5,7,10,12\}$; $\ell=1$ is top
left and $\ell=12$ is bottom-right.}\label{fig11}
\end{figure}

\newpage
\begin{figure}
\begin{center}
\includegraphics[width=6in]{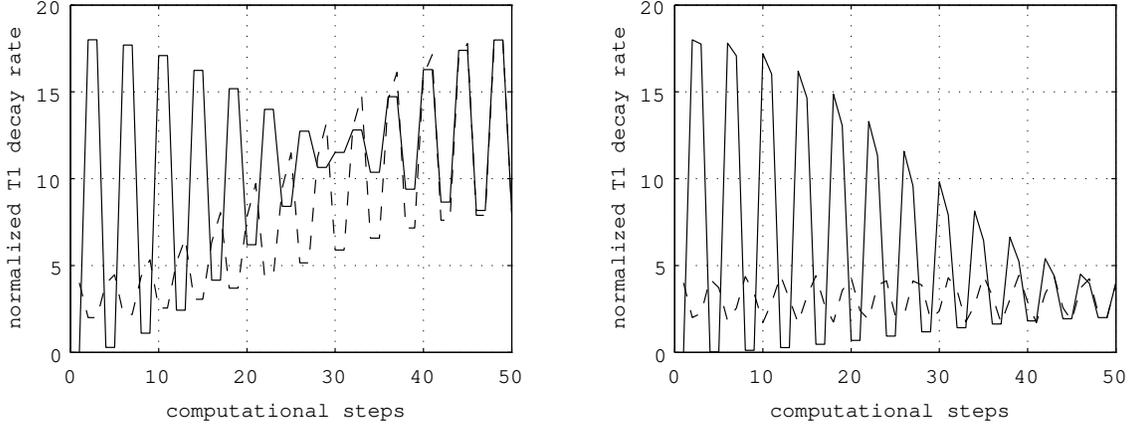}
\caption{The normalized $T_1$ relaxation rates at each of the 50
computational steps for the $n=8$ database search example; the left
figure applies to the target state $\left|\tau=255\right>$ and the
right figure is for $\left|\tau=15\right>$. The first Hadamard gate
is followed by twelve Grover iterations consisting of four steps of
$-\hat{H}\hat{I}_0\hat{H}\hat{I}_{\tau}$. The solid and dashed lines
represent results for the standard implementation using an initial
state $|\psi _1 \rangle =|0\rangle $ and the improved implementation
using an initial state $|\psi _1 \rangle =|15\rangle
=|----++++\rangle$, respectively.} \label{fig12}
\end{center}
\end{figure}

\begin{figure}
\begin{center}
\includegraphics[width=6in]{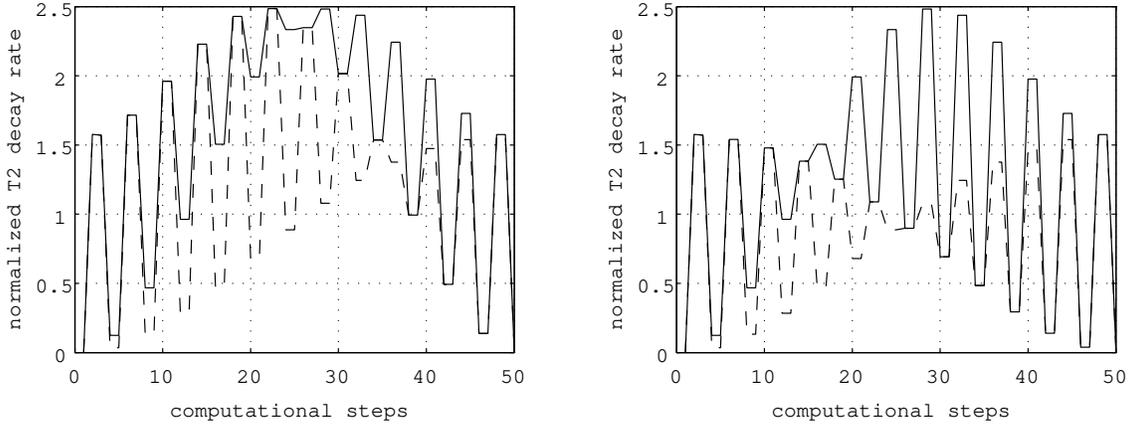}
\caption{The normalized $T_2$ relaxation rates for the $n=8$ search
algorithm; plots are labeled as in Fig. \ref{fig12}.} \label{fig13}
\end{center}
\end{figure}


\begin{figure}
\includegraphics[width=4in]{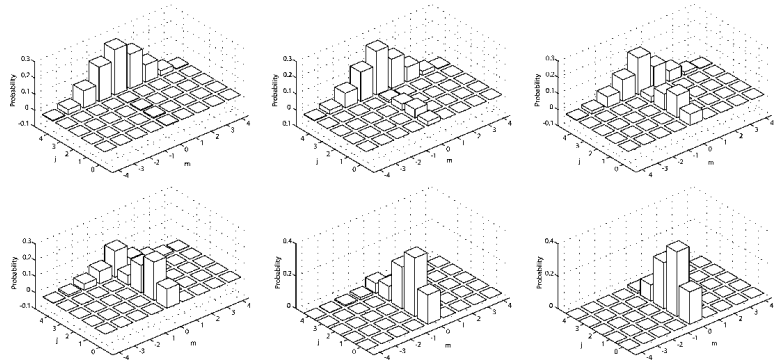}
\caption{$P(j,m)$ of the control register states for the second
$n=8$ example of the database search algorithm. The initial and
final states are $|\psi _1\rangle = |0\rangle $ and $|\psi _4\rangle
= |15\rangle $.} \label{fig14}
\end{figure}
\clearpage

\end{widetext}
\end{document}